\documentclass[12pt]{article}
\usepackage{epsf}
\usepackage{amsmath}

\usepackage{graphics}
\usepackage{cite}

\begin{titlepage}

\begin{document}

\hfill October 2022

\begin{center}

{\bf \LARGE Electromagnetic Accelerating Universe}\\
\vspace{2.5cm}
{\bf Paul H. Frampton}\footnote{paul.h.frampton@gmail.com}\\
\vspace{0.5cm}
{\it Dipartimento di Matematica e Fisica "Ennio De Giorgi",\\ 
Universit\`{a} del Salento and INFN-Lecce,\\ Via Arnesano, 73100 Lecce, Italy.
}

\vspace{1.0in}

\begin{abstract}
\noindent
In recent work we have extended the theory that dark matter is composed
of primordial black hole (PBHs) to extremely high masses and made an
assumption that the holographic entropy bound is saturated.
Astrophysicists
have recently suggested that PBHs are formed with electric charges Q,
retain their charges for the age of the universe, 
all charges have the same sign, and Q/M increases with mass M. Adopting these
assumptions and using an approximate formula relating Q and M, it
is here suggested that, for PBHs with over a trillion solar masses, Coulomb
repulsion can exceed gravitational attraction and that electromagnetic
properties of dark matter are a possible cause for the observed accelerated cosmological
expansion.
\end{abstract}

\end{center}
\end{titlepage}

\section{Introduction} 
\bigskip
The idea that the constituents of dark matter are primordial
black holes (PBHs) is gaining momentum. Using entropy
arguments, and assuming that the holographic entropy
bound is saturated by PBHs led us in \cite{Frampton,Frampton2}
to introduce four tiers of PBHs according to their mass.

\bigskip

\noindent 
Adopting as the unit of mass the solar mass $M_{\odot} = 2\times 10^{30}$kg.
we define a mass exponent $p$ by $M_{PBH} \equiv 10^p M_{\odot}$. In terms
of $p$, the four tiers are defined with their corresponding
acronyms as follows:

\begin{itemize}
\item Tier 1:  PBHs with $p<2$.
\item Tier 2:  PIMBHs with $2<p<5$ ~~ (IM $\equiv$ Intermediate Mass)
\item Tier 3:  PSMBHs with $5<p<11$ ~~ (SM $\equiv$ Supermassive)
\item Tier 4:  PEMBHs with $11<p<22$ ~~ (EM $\equiv$ Extremely Massive)
\end{itemize}

\bigskip

\noindent
In the first several decades after the first proposal of PBHs in the 1960s,
it was assumed that they were all in Tier 1. In fact it was assumed that
$p \ll 0$ and that PBHs were not only much lighter than the Sun but also
than the Earth or even than asteroids.

\bigskip

\noindent
The idea that PBHs may be formed with higher masses than Tier 1 arose
in the last few decades. For example, in 2010\cite{FKTY} it was shown that, at
least mathematically, PBHs of arbitrarily large mass exponent $p$ could be
formed from arbitrarily large fluctuations and inhomogeneities. 

\bigskip

\noindent
Tier 2 PIMBHs with $2<p<5$ were suggested\cite{Frampton2015,ChaplineFrampton} in 2015 as good candidates
for the dark matter within galaxies such as the Milky Way. Within the Milky Way, a promising
method for detection of PIMBHs is by microlensing, using as targets the stars in the
Magallenic Clouds. At the time of writing, all attempts to find the relevant multi-year-duration
microlensing light curves have proved to be inconclusive.

\bigskip

\noindent
Tier 3 PSMBHs with $5<p<11$ include the supermassive black holes located
at galactic centres. Of the known contents of the universe, these dominate by far
the entropy of the universe by 15 orders of magnitude but still fall far
short of the holographic entropy bound by another 20 orders of
magnitude. This last was the motivation for introducing Tier 4.

\bigskip

\noindent
Tier 4 PEMBHs with $11<p<22$ were the  suggestion made in
\cite{Frampton,Frampton2} as the only known way to saturate
the holographic entropy bound. We cut off
Tier 4 at $p=22$ only because $p=23$ represents the total mass of the
visible universe. A PBH with $p=22$ may seem unlikely, but the dark matter
is sufficiently mysterious that it is best to keep all possibilities open.
 
 \bigskip
 
 \noindent
 In the present paper we suggest
a second argument to support the existence of Tier 4
which is related to  
the observed accelerated cosmological expansion.

\newpage

\section{Electrically charged PBHs}

\noindent
Kerr black holes are completely characterised by three
parameters mass M, electric charge Q and spin S. In
our previous discussions we have ignored S although
it is to be expected that for a PEMBH the extremely
high mass will be generally accompanied by a
very large spin angular momentum.

\bigskip

\noindent
We have tacitly assumed that $Q=0$ as is normally
done for astrophysical objects. This assumption has
been recently queried for PBHs in \cite{Araya} (for
earlier related papers, see \cite{Komissarov,KingPringle,Zajacek}).

\bigskip

\noindent
In \cite{Araya}, four questions are addressed:

\begin{itemize}
\item 
(1) Are PBHs formed with non-zero electric charge Q?
\item 
(2) Do PBHs retain this charge for at least the age of the universe?
\item
(3) Do the PBH charges all have the same sign?
\item
(4) Does the ratio Q/M of PBHs increase with increasing mass M?
\end{itemize}

\bigskip

\noindent
Remarkably, the authors arrive at positive answers for all of these 
four questions, and these answers will be assumed in the following.
For question (3) a common negative charge like the electron is
favoured, although in what follows the sign of the PBH charges
does not matter, as long as it is common. We shall focus on the
very interesting positive answer to question (4).

\bigskip

\noindent
The positive answer to (3) is crucial to our cosmological model
and its underlying reason is explained as follows.
After recombination at $Z\sim 1100$ there are no free charges
to be accreted by PBHs. Later, re-ionisation by ultra-violet
emission from the first stars leads to separation of electrons
from atomic nuclei and both achieve thermal equilibrium.
Because the electrons are lighter, their thermal velocities
are higher and this strongly favours their accretion yielding
PBHs all with the same sign of electric charge {\it viz.}
negative.

\bigskip

\noindent
It should be noted that in the present model the {\it only}
large sized objects with significant electric
charge are PBHs with extremely high mass. Other matter,
baryonic or dark, is electrically neutral and the origin of the
large scale structure is not changed.

\bigskip

\noindent
We shall define a charge exponent $q$ for PBHs by their
charge Q being $Q=10^q$ Coulombs. The relationship
between $q$ and the mass exponent $p$ will be crucial
and to suggest what this is we need to extrapolate
the results in \cite {Araya} about question (4) 
outside of their range of validity, to be justified only
by our interesting conclusions.

\bigskip

\noindent
Let us consider two identical PBHs both with mass $M=10^pM_{\odot}$
and electric charge $Q=10^q$C. Let the magnitudes of the gravitational
attraction and Coulomb repulsion be $F_g$ and $F_E$ respectively.
The ratio of these will be denoted by
\begin{equation}
{\cal R} \equiv \left(\frac{F_g}{F_E} \right).
\end{equation}
which scales as $(M^2/Q^2) \equiv 10^{2(p-q)}$.

\bigskip

\noindent
To calculate $({\cal R})_{PBH}$, we may start from the well-known
fact that for the electron and proton in a hydrogen atom,
\begin{equation}
\left( {\cal R} \right)_{H-atom} = 10^{-39}
\label{Hatom}
\end{equation}
and then scale by $(M^2/Q^2)$ to obtain
\begin{equation}
\left( {\cal R} \right)_{PBH} = 10^{-39} \left( \frac{M_{PBH}^2}{M_eM_p} \right)
\left( \frac{e}{Q_{PBH}} \right)^2
\label{Rpbh}
\end{equation}

\bigskip

\noindent
Changing mass units to solar masses $M_{\odot}$, Eq(\ref{Rpbh}) becomes
\begin{equation}
\left( {\cal R} \right)_{PBH} = 6.8 \times 10^{40 + 2(p-q)}
\label{Rpbh2}
\end{equation}

\bigskip

\noindent
To pass from Eq.(\ref{Rpbh}) to Eq.(\ref{Rpbh2}) we used
\begin{itemize}
\item
$M_eM_p = 470 MeV^2$
\item
$1 MeV^2 = 3.2 \times 10^{-60} kg^2$
\item
$1 kg^2 = 2.5 \times 10^{-61} M_{\odot}^2$
\item
$e^2 = 2.56 \times 10^{-38} C^2$
\end{itemize}

\noindent
To use Eq.(\ref{Rpbh2}) we need a relationship between
the mass exponent $p$ and the charge exponent $q$.
For this we extrapolate from the following results in \cite{Araya}

\begin{itemize}
\item
$Q/M = 10^{-32}$ C/kg for $M=10^{20}$ kg.
\item
$Q/M = 10^{-22}$ C/kg for $M=10^{30}$ kg.
\end{itemize}

\bigskip

\noindent
These results suggest an increase in $Q/M$ with increasing $M$.
It is irresistible to attempt an extrapolation beyond their domain
of validity as follows

\begin{itemize}
\item
$Q/M = 10^{m-52}$ C/kg for $M=10^m$ kg.
\end{itemize}

\bigskip

\noindent
This extrapolation to higher mass PBHs leads to the following
relationship between $q$ and $p$

\begin{equation}
q = 8 + 2p + \log 4
\label{qp}
\end{equation}
Using Eq.(\ref{qp}) in Eq.(\ref{Rpbh2}), we display in Table 1
some examples of Tier 2, 3, 4 PBHs with $2 \le p \le 20$.

\bigskip

\noindent
The values displayed in Table 1 reveal the remarkable fact that as the PBH mass
increases into the Tier 4 of PEMBHs introduced in \cite{Frampton,Frampton2}
the Coulomb repulsion begins to exceed the gravitational attraction somewhere
between $p=11$ and $p=14$. 

\bigskip

\noindent
We can be more specific by solving ${\cal R} = 1$ and find $p=11.8$. This implies
that in the case of two PEMBHs each with, for example, one trillion solar masses the
Coulomb repulsion is of comparable order of magnitude to the gravitational
attraction and electromagnetic repulsion dominates gravitational attraction 
as the PEMBH mass further increases. This is counterintuitive 
only because astrophysical
objects are usually assumed to carry negligible electric charge.

\bigskip

\begin{table}[h]
\caption{Values of ${\cal R}$ and ${\cal R}^{-1}$  for PBHs with $M_{PBH} = 10^p M_{\odot}$  for various $p$.
The value $q=(8+2p+log4)$ is used.}
\begin{center}
\begin{tabular}{||c|c|c|c|c||}
\hline
MASS & CHARGE  & ${\cal R} = F_g/F_E$ & ${\cal R}^{-1}=F_E/F_g$   \\
 $10^p M_{\odot}$ & $10^q$ C & =$6.8\times10^{40+2(p-q)}$ & \\
\hline
\hline
p=2 & $q=12+\log 4$ & $4.2 \times 10^{19}$ & $2.4 \times 10^{-20}$   \\
\hline
p=5 & $q=18+log4$ & $4.2 \times 10^{13}$ & $2.4 \times 10^{-14}$  \\
\hline
p=8 & $q=24+log4$ & $4.2\times 10^7$ & $2.4\times 10^{-8}$  \\
\hline
p=11& $q=30+log4$  & 42 & $2.4\times10^{-2}$ \\
\hline
p=14& $q=36+log4$ & $4.2\times 10^{-5}$ & $2.4\times 10^4$  \\
\hline
p=17& $q=42+log4$ & $4.2\times 10^{-11}$ & $2.4\times 10^{10}$ \\
\hline
p=20& $q=48+log4$ & $4.2\times10^{-17}$ & $2.4\times10^{16}$  \\
\hline
\end{tabular}
\end{center}
\label{longevity}
\end{table}

\bigskip

\noindent
Electric charges for PEMBHs are large. Following from
our formulas Eqs. (\ref{Rpbh2}) and (\ref{qp}) we list for a few notable cases
what are the electric charges in Coulombs, together with the corresponding ratio (${\cal R}^{-1}$)
of the  electromagnetic repulsion to the gravitational attraction for two identical
PEMBHs.
\begin{itemize}
\item 
p=12 (one trillion  solar masses) has $Q=4\times 10^{32}$ C \\and ${\cal R}^{-1}=2.2$.
\item
p=18 (estimated mass of the Great Attractor) has $Q= 4\times 10^{44}$ C \\and ${\cal R}^{-1} = 2.2\times 10^{12}$.
\item
p=22 (the highest mass PEMBH considered) has $Q= 4\times 10^{52}$ C\\
and ${\cal R}^{-1}= 2.2\times 10^{20}$ 
\end{itemize}

\bigskip

\noindent
It has not escaped our attention that the electric repulsion between PEMBHS could
provide an explanation for the observed cosmological acceleration.

\newpage

\section{Discussion}

\noindent
When the accelerating expansion of the universe was discovered in 1998
\cite{Riess,Perlmutter}, it surprised everyone. It  was a monumental
discovery which was counterintuitive.
To provide a theoretical explanation for this discovery requires a comparably
monumental explanation and any specific theory has an extremely small
chance of being correct. We suspect that the present theory is no
exception.

\bigskip

\noindent
That being said, the present theory does have some redeeming qualities. The discovery
of accelerated expansion was counterintuitive because the force of gravity is
always attractive. Our present theory is counterintuitive because electromagnetic forces
between astrophysical objects are generally infinitesimally small and 
negligible in astronomy and cosmology.

\bigskip

\noindent
In the hydrogen atom the gravitational attraction between
the proton and electron is completely negligible compared
to the electromagnetic attraction. By contrast, in the Solar System the
gravitational attraction dominates because the electric charges
of the Sun and planets are negligibly small and the same is true
for galaxies and clusters of galaxies.

\bigskip

\noindent
In the present theory, the dark matter PBHs have extraordinary
electromagnetic properties. For the
first three tiers of PBHs, relevant to galaxies and clusters, gravitational attraction 
continues to dominate. For Tier 4 PBHs, however, which constitute the
intergalactic and cosmological dark matter discussed in \cite{Frampton,Frampton2},
as the PBH mass increases there is a tipping point at about one trillion
solar masses beyond which, because of electric charges with a common sign,
electromagnetic repulsion exceeds gravitational attraction. It is this electromagnetic
repulsion
which could explain the accelerated expansion of the universe.

\bigskip

\noindent
It will be interesting to discover how the present theory stands up to more
assiduous and sedulous study.

\newpage

\section*{Acknowledgement}

\noindent
We thank the Department of Physics at UniSalento for an affiliation.

\end{document}